\documentclass[a4paper,11pt]{article}

\usepackage{jcappub} 

\usepackage{xcolor}
\usepackage{bm}
\usepackage{ulem}
\usepackage{graphicx}
\usepackage{geometry}
\usepackage{multirow}
\usepackage{amsmath}
\usepackage{amssymb}
\usepackage{mathtools}
\usepackage{blindtext}

\usepackage{babel}

\title{\boldmath{The effective field theory approach to the strong coupling issue in $f(T)$ gravity with a non-minimally coupled scalar field}}

\author[a,b]{Yu-Min Hu, }
\author[c]{Yang Yu, }
\author[a,b,1]{Yi-Fu Cai, }
\author[c,1]{Xian Gao\note{Corresponding authors.}}

\affiliation[a]{Deep Space Exploration Laboratory/School of  Physical Sciences, University of Science and Technology of China, Hefei, Anhui 230026, China}
\affiliation[b]{Department of Astronomy, School of Astronomy and Space Science, University of Science and Technology of China, Hefei, Anhui 230026, China}
\affiliation[c]{School of Physics and Astronomy, Sun Yat-sen University, Zhuhai 519082, China}

\emailAdd{yumin28@ustc.edu.cn}
\emailAdd{yuyang69@mail2.sysu.edu.cn}
\emailAdd{yifucai@ustc.edu.cn}
\emailAdd{gaoxian@mail.sysu.edu.cn}

\abstract{The Hamiltonian analysis for $f(T)$ gravity implies the existence of at least one scalar-type degree of freedom (DoF). However, this scalar DoF of $f(T)$ gravity does not manifest in linear perturbations around a cosmological background, which indicates an underlying strong coupling problem. In this work we expand the scope by introducing an extra scalar field non-minimally coupled to $f(T)$ gravity, aiming to address or alleviate the aforementioned strong coupling problem. Employing the effective field theory (EFT) approach, we provide a class of torsional EFT forms up to second order operators, avoiding the Ostrogradsky ghost. To illustrate this phenomenon, we study a simple model and perform a detailed analysis of its linear scalar perturbations. The results demonstrate that the coupling terms in this toy model are necessary to avoid the initial degenerate situation. The complete avoidance of new constraints requires more coupling terms. Once this vanishing scalar DoF starts propagating in cosmological background at linear level, this phenomenon will demand a revisit of the strong coupling issue that arises in $f(T)$ gravity, particularly in the presence of matter coupling. }

\begin{document} 

\begin{sloppypar}
\maketitle
\flushbottom

\section{Introduction}

In recent years, interpretations of the spacetime geometry other than the Riemannian description have illuminated different perspectives for understanding the nature of gravity \cite{Boehmer:2021aji, CANTATA:2021ktz}. 
Teleparallelism, in which the spacetime curvature is supposed to be vanishing for absolute parallel, has attracted much attention. 
The so-called ``trinity of gravity'' has been proposed \cite{BeltranJimenez:2019esp, Heisenberg:2018vsk, Capozziello:2022zzh} (see \cite{Bahamonde:2021gfp} for recent reviews), in which the inherent relation among torsion, non-metricity and curvature has been revealed. 
In the absence of curvature and non-metricity, a theory equivalent to general relativity (GR) is described by the Lagrangian that is linear in the torsion scalar, which is dubbed the teleparallel equivalent of general relativity (TEGR) \cite{Maluf:2013gaa, Aldrovandi:2013wha}. 
In fact, TEGR can be traced back to the discussion proposed by Einstein himself in the 1920s \cite{Goenner:2014mka}. 
Similarly, the last piece of jigsaw for the whole trinity is formulated by the non-metricity without curvature or torsion.
The Lagrangian is linear in the non-metricity scalar, which is dubbed the symmetric teleparallel equivalent of general relativity (STEGR) \cite{Nester:1998mp}. 
Although the torsion and the non-metricity scalars are equivalent to GR, when non-linear extensions such as $f(T)$ modified gravity are taken into account \cite{Cai:2015emx, Krssak:2015oua, Krssak:2018ywd, Hohmann:2019nat, Bahamonde:2017wwk, Jarv:2018bgs, Heisenberg:2023lru}, the dynamics could be drastically different from that of GR. 
Many efforts have been devoted to the investigation of the cosmological implications of these theories \cite{Zheng:2010am, Bengochea:2010sg, Tamanini:2012hg, Cardone:2012xq, Farrugia:2016xcw, Chen:2019ftv, Cai:2019bdh, Golovnev:2020aon, Ren:2021uqb,Fiorini:2021mps, Ren:2021tfi, Golovnev:2021htv, BeltranJimenez:2021kpj, Duchaniya:2022rqu, DeBenedictis:2022sja, Zhao:2022gxl, Huang:2022slc, dosSantos:2021owt, Chen:2022wtz, Zhang:2023scq, Heisenberg:2023wgk, Heisenberg:2023tho}. 

Teleparallel theories of modified gravity usually exhibit distinct features when compared with GR or more general curvature-based gravity. 
When considering the number of degrees of freedom (DoFs), a Hamiltonian analysis is necessary, though complicated. Generally,  results presented in the literature \cite{Blixt:2018znp, Blixt:2019ene, Guzman:2020kgh, Golovnev:2021omn, Hu:2022anq} support the existence of extra DoFs and reveal the non-constant rank of the Poisson brackets' algebra. 
It is commonly accepted that $f(T)$ theory propagates at most 5 DoFs \cite{Li:2011rn, Blagojevic:2020dyq} (see \cite{Blixt:2020ekl} for a review), including an additional scalar-type DoF. However, with different field configurations, the number and algebra of constraints may change, leading to a direct reduction in the rank of Poisson brackets and a corresponding decrease of the number of DoFs \cite{Chen:2014qtl}. 

To address this uncertainty, perturbation methods offer valuable insights, especially when applied to a specific background. 
However, in the context of teleparallel modified gravity, the underlying strong coupling issues, which are indicated by the absence of dynamical modes in linear perturbation, have been subsequently demonstrated as a general feature \cite{BeltranJimenez:2019nns, Golovnev:2020nln, BeltranJimenez:2021auj}.
Around a Minkowski background, an extra dynamical scalar mode of $f(T)$ gravity manifests itself only in the fourth order perturbation action \cite{BeltranJimenez:2020fvy}. 
The absence of the scalar mode in lower-order perturbations raises concerns about a strong coupling problem \cite{Blas:2009ck, Papazoglou:2009fj, Bueno:2016xff, BeltranJimenez:2019tme, BeltranJimenez:2020lee, Hell:2021oea, Hell:2023mph}, which can be understood from an effective point of view---if the kinetic term is considered to be small, normalizing the kinetic term by redefining the variables will inevitably induce and amplify the coupling coefficients at higher order in perturbations. 
Linear perturbations of $f(T)$ gravity in a Friedmann-Lema\^\i tre-Robertson-Walker (FLRW) universe have been explored in previous studies \cite{Dent:2010nbw, Chen:2010va, Izumi:2012qj, Golovnev:2018wbh, Sahlu:2019bug, Bahamonde:2022ohm, Hohmann:2020vcv}, revealing no additional DoFs at linear level. In order to quantify the effects of the strong coupling, second order perturbations of $f(T)$ gravity are studied in a flat FLRW background and an estimation of strong coupling scale is performed as well by effective field theory (EFT) method \cite{Hu:2023juh}. 

This study concentrates on torsional geometry, aiming to reexamine the strong coupling issue within the context of $f(T)$ gravity, incorporating an additional non-minimally coupled scalar field.
Numerous investigations have explored various types of matter coupling \cite{Arcos:2004tzt, Harko:2014sja, BeltranJimenez:2020sih, Heisenberg:2022mbo, Harko:2018gxr} in teleparallel gravity, especially involving a scalar field \cite{Skugoreva:2014ena, Fazlpour:2014qla, Otalora:2014aoa, Hohmann:2018dqh, Bahamonde:2019shr}. 
As the first step, we consider a class of abstract action including non-trival coupling terms without Ostrogradsky ghost and express it in EFT form up to second order operators as a representative generalization in the torsion-based EFT framework \cite{Li:2018ixg, Cai:2018rzd}. 
It is worth noting that the EFT approach has been applied to inflationary cosmology \cite{Creminelli:2006xe, Cheung:2007st, Ashoorioon:2018uey}, and then to dark energy \cite{Creminelli:2008wc, Gubitosi:2012hu, Bloomfield:2012ff}, as well as to torsion gravity from observational perspectives \cite{Yan:2019gbw, Yan:2019hxx, Ren:2022aeo}. 
The EFT approach naturally cooperates and characterizes the extra scalar field with spacetime geometry \cite{Arkani-Hamed:2003pdi, Gleyzes:2013ooa}.
Moreover, it is also a tool for dealing with cosmological perturbation since it allows one to investigate the background and perturbations at various orders separately \cite{Piazza:2013coa}. 

This paper is organized as follows. In Sec. \ref{sec:2}, an illustration for torsional teleparallel gravity is given. In Sec. \ref{sec:3}, we briefly review the EFT approach to torsional gravity and further bring in specific second order operators for our consideration. The new operators systematically introduced will be used for future discussions on the scalar DoF of $f(T)$ theory and hopefully redefining the strong coupling problem. 
In Sec. \ref{sec:4}, a simple torsional EFT model up to second order is presented. We proceed with a linear scalar perturbations analysis in cosmological background and perform a rough constraints analysis at the Lagrangian level. This is aimed at suggesting the potential to avoid the strong coupling issue in $f(T)$ gravity. Our results are summarized in Sec. \ref{sec:5}. Throughout this paper we choose the unit $8\pi G=M_P^2$ with the Planck mass $M_P$ and the convention for the metric $\{-,+,+,+\}$.

\section{Teleparallel gravity and $f(T)$ gravity}\label{sec:2}

In this section, we briefly review the fundamental conceptions of teleparallel gravity. Teleparallel geometry is derived from the metric-affine theory with vanishing curvature, in which the torsion tensor $T^{\lambda}{}_{\nu\mu}$ is not negligible. Meanwhile the notion of tetrad fields $e^{\phantom{A}\mu}_{A}$ is introduced as an alternative description, where Greek indices stand for spacetime coordinates and Latin indices correspond to tangent space coordinates. 
To be specific, an orthonormal basis $e_{A} $ for the tangent space at each point $x^\mu$ of the 4-dimensional manifold can be related to the natural basis through
\begin{equation} 
     e_{A}\coloneqq e^{\phantom{A}\mu}_A\partial_{\mu}, \qquad e^{A}\coloneqq e^A_{\phantom{A}\mu} {\rm d}x^{\mu}~.
\end{equation} 
The co-tetrads $e_{\phantom{A}\mu}^A$ satisfy $e^{\phantom{A}\mu}_A e^A_{\phantom{A}\nu} = \delta^{\mu}_{\nu}$ and $e^{\phantom{A}\mu}_A e^B_{\phantom{A}\mu} = \delta^A_B$, and are related to the metric through $g_{\mu \nu}=\eta_{A B} e^{A}{}_{\mu} e^{B}{}_{\nu}$, with the Minkowski metric $\eta_{AB}=\text{diag}\{-,+,+,+\}$. 

For the affine-connection $\Gamma^\lambda{ }_{\nu \mu}$, the tetrad postulate implies its relation with the spin connection
\begin{align}
\hat\nabla_{\mu}e_{\ \nu}^{A}=\partial_{\mu}e_{\ \nu}^{A}-\Gamma_{\ \nu\mu}^{\lambda}e_{\ \lambda}^{A}+\omega_{\ B\mu}^{A}e_{\ \nu}^{B}=0~,
\end{align}
where $\hat\nabla_{\nu}$ is the covariant derivative associated with the affine connection. 
The spin connection is used to describe inertial effects representing pure gauge in flat background and transforms under Lorentz transformation $\Lambda$ as
\begin{equation}
\omega^{A}_{\ B\mu}\rightarrow \Lambda^{A}_{\ C}(\Lambda^{-1})^{D}_{\ B}\omega^{C}_{\ D\mu}+\Lambda^{A}_{\ C}\partial_{\mu}(\Lambda^{-1})^{C}_{\ B}~.
\end{equation}
In this case we are able to choose one special set of tetrad fields such that the spin connection vanishes identically, $\omega^{B}_{\phantom{B}A\mu}\equiv 0$, which is dubbed the Weitzenb\"{o}ck gauge. Under this gauge, the affine-connection transforms to
\begin{align}
\Gamma^\lambda{ }_{\nu \mu}=e_A{}^\lambda \partial_\mu e^A{ }_\nu~,
\end{align}
which is the so-called Weitzenb\"{o}ck connection and depends on the tetrads only with vanishing curvature. Correspondingly, the torsion tensor is the antisymmetric part of the affine-connection
\begin{align}
{T}^{\lambda}{}_{\mu \nu}=\Gamma^\lambda{ }_{\nu \mu}-\Gamma^\lambda{ }_{\mu 
\nu}=e_{A}{}^{\lambda}(\partial_{\mu} e^{A}{}_{ 
\nu}-\partial_{\nu} e^{A}{}_{\mu})~.
\end{align}
Moreover, the contortion tensor represents the difference between the Levi-Civita connection $\mathring{\Gamma}^{\rho}{}_{\mu \nu}$ in Riemannian spacetime and Weitzenb\"{o}ck connection $\Gamma^\rho{ }_{\mu \nu}$ in Weitzenb\"{o}ck spacetime,
\begin{equation}
\label{eq:conrelation}
	\mathcal{K}^{\rho}{}_{\mu \nu}={\Gamma}^{\rho}{}_{\mu \nu}-\mathring{\Gamma}^{\rho}{}_{\mu \nu} ~.
\end{equation}
The contortion tensor can also be reformulated into a combination of torsion tensors, namely
\begin{align}
\mathcal{K}^{\rho}{}_{\mu \nu} \equiv \frac{1}{2} \Big( T_{\mu}{}^{\rho}{}_{\nu}+{T_{\nu}{}^{\rho}}_{\mu} -{T^{\rho}}_{\mu \nu} \Big)~.
\end{align}
With the vanishing curvature, teleparallel gravity gives an equivalence between the Ricci scalar $R$ corresponding to the Levi-Civita connection and the torsion scalar $T$ up to a boundary term \cite{Aldrovandi:2013wha, Maluf:2013gaa}
\begin{equation}
R=-T-2 \nabla_\mu T^\mu~,
\label{eq:relationRT}
\end{equation}
where the torsion scalar is a contraction of the torsion tensors
\begin{equation}
    \label{eq:Tscalardef}
    T = S_{\rho}{}^{\mu \nu} T^{\rho }{}_{\mu \nu}=\frac{1}{4} T^{\rho}{}_{\mu 
\nu} T_{\rho}{}^{\mu \nu}+\frac{1}{2} T^{\rho}{}_{\mu \nu} T^{\nu 
\mu}{}_{\rho}-T^{\rho}{}_{\mu \rho} T^{\nu \mu}{}_{\nu} ~,
\end{equation}
with the super-potential $
	S_{\rho}{}^{\mu \nu} \equiv \frac{1}{2} \Big( {K}^{\mu\nu}{}_{\rho} 
+\delta_{\rho}^{\mu} T^{\alpha \nu}{}_{\alpha} -\delta_{\rho}^{\nu} T^{\alpha 
\mu}{}_{\alpha} \Big)$, and the torsion vector appearing within the boundary term is defined by
\begin{align}
T^\mu\equiv T^{\nu \mu  }{}_{\nu}~.
\end{align}

Inspired by the  $f(R)$ modifications of gravity, one can proceed to the construction of modified teleparallel gravity by promoting the torsion scalar in the action to a function $f(T)$, resulting to $f(T)$  gravity, namely
\begin{equation}
 S = \int d^{4} x\, e \,\frac{M_{P}^{2}}{2} f({T}) ~,
\end{equation}
where the torsion scalar $T$ is constructed by the Weitzenb\"{o}ck connection, $e=\det{(e{^A}{_\mu})=\sqrt{-g}}$ and $M_{P}$ is the Planck mass. 
In this work, we are interested in the strong coupling issue in $f(T)$ gravity. 
The action considered in this paper is given by 
\begin{equation} \label{action}
 S = \int d^{4} x\, e \,\Big[\frac{M_{P}^{2}}{2} f({T}) + \mathcal{L}_{\text{scalar}}+\mathcal{L}_{\text{couple}}\Big] ~,
\end{equation}
in which an extra scalar field coupled with torsional gravity is taken into account.
To specify what is concerned, we explain the action form \eqref{action} and identify the relevant DoFs within the scope of this paper. 
First, from the perspective of pure gravity,
\begin{itemize}
    \item we exclusively consider a functional form of the torsion scalar, namely, $f(T)$ gravity.
\end{itemize}
Next, we have to answer the question that up to which order in derivatives of the scalar field are allowed, since higher order derivative terms such as  $\hat\nabla_{\mu} \hat\nabla_{\nu}\Phi$ usually give rise to undesired ghost instabilities.
For the sake of simplicity, we require that 
\begin{itemize}
    \item  only derivatives of the scalar field up to the second order are allowed, and should enter in the form of the commutator $\hat\nabla_{[\mu} \hat\nabla_{\nu]}\Phi$.
\end{itemize}
If $\hat{\nabla}_{\mu}\hat{\nabla}_{\nu}\Phi$ enters in the action in the form of the commutator, higher order time derivatives will be canceled without any additional condition. For simplicity and ghost-free requirement, pairs of commutators $\hat\nabla_{[\mu} \hat\nabla_{\nu]}\Phi$ are the only allowed higher derivatives in the action.
Then the form of $\mathcal{L}_{\text{scalar}}$ is restricted to be a simple form 
\begin{align} \label{action scalar}
    \mathcal{L}_{\text{scalar}}=\mathcal{L}_{\text{scalar}}(g^{\mu\nu};\Phi,\hat\nabla_\mu \Phi, \hat\nabla_{[\mu} \hat\nabla_{\nu]}\Phi)~.
\end{align}
These building blocks can enter in $\mathcal{L}_{\text{scalar}}$ in any form of scalar contractions, which prevent $\mathcal{L}_{\text{scalar}}$ from the appearance of Ostrogradsky ghost. 
Regarding the non-minimal coupling part $\mathcal{L}_{\text{couple}}$, we do not consider any index contraction between torsion tensors since some of them will appear in our following second order EFT action, which would make our argument ambiguous with respect to other modified torsional theories.
Typically these theories exhibit more complex behavior in terms of degrees of freedom.
In practice, our purpose is to investigate the potential of using coupling terms to address the issue of strong coupling in $f(T)$ theory. Therefore,
\begin{itemize}
    \item only the contractions between torsion and derivative of the scalar field are considered.
\end{itemize}
As a result, $\mathcal{L}_{\text{couple}}$ is given by 
\begin{align} \label{action coupling}
    \mathcal{L}_{\text{couple}}=\mathcal{L}_{\text{couple}}(T,T^\mu,T^{\mu \nu \rho} ;\Phi, X, \hat \nabla_\mu \Phi, \hat\nabla_{[\mu} \hat\nabla_{\nu]}\Phi)~,
\end{align}
where $X$ is defined as $X=-\frac{1}{2}\hat\nabla_{\mu}\Phi \hat\nabla^{\mu}\Phi$.

In this work, a given action \eqref{action} satisfying all these requirements will consequently contain two scalar type DoFs — namely, one extra scalar DoF in $f(T)$ gravity and the other from the additional scalar field.
We have already known that none of the two scalar DoFs will always be present as dynamical modes in linear scalar perturbation around a given FLRW cosmological background, which is a characteristic signal of strong coupling problem in a perturbation theory.
However, we expect the possibility that the non-minimal coupling terms may play a role in changing the constraint structure and avoiding, or at least alleviating the strong coupling issue in $f(T)$ gravity. 
Interestingly, the linear boundary term, defined as $B=-2 \nabla_\mu T^\mu$, with coupling to $\Phi$ is degenerate with the coupling term $T^\mu \nabla_{\mu}\Phi$ through a integration by parts. Nonlinear higher-order derivative terms from both sides will increase the order of the equations of motion in general, and theories typically come with unwanted ghost-like DoFs, making the problem more complex.

\section{The EFT approach of $f(T)$ gravity coupled with a scalar field}\label{sec:3}

We will apply the EFT approach and rewrite the action \eqref{action}, namely $f(T)$ gravity coupled with a scalar field in the previous section, which is regarded as a low-energy effective theory around the FLRW background.
In short, we do not restrict ourselves to the case of pure $f(T)$ gravity, where both matter component and non-minimal coupling terms are neglected in the EFT action. 
We emphasize that in this work we do not provide the most general form of a ghost-free EFT action up to the second order operators.
The corresponding EFT action is within the framework based on our requirements and would be used to study its linear scalar perturbations and possible strong coupling issue.

Generally, we introduce an additional time-like dynamical scalar field $\Phi$ with its decomposition in a perturbed FLRW Universe 
\begin{equation}
 \Phi(t,\vec{x}) = \Phi_0(t) + \delta\Phi(t,\vec{x}) ~,
\end{equation}
where $ \Phi_0$ is the homogeneous background value of the scalar field and $\delta \Phi$ is its perturbation.   
Then, under the unitary gauge, $\delta \Phi$ is fixed to zero with the gradient of the scalar field expressed as
\begin{align}
    \hat{\nabla}_{\mu}\Phi=\nabla_{\mu}\Phi\overset{\text{u}}{=}\dot \Phi_0 \delta^0_\mu~.
\end{align}
where the dot(s) symbol over variables is denoted as temporal derivative here and in the following, e.g. $\dot \Phi_0= \partial_t \Phi_0$ and the number of dots stands for the order of derivative. Time derivative $\dot \Phi_0$ is regarded as a background value and would be absorbed in the coefficient functions. 
For simplicity, in the flat FLRW background, we usually normalize the background value of $g^{00}$ and set $g^{00}=-1+\delta g^{00}$.
It is easy to see that a linear kinetic term $\nabla_\mu \Phi \nabla^\mu \Phi$ contributes to the operator $g^{00}$.

In principle, one should treat both dynamical variables, namely, the tetrad fields and the scalar field, on an equal footing within the EFT framework, since all these fields are incorporated into the torsional EFT action through fundamental geometric operators that respect the symmetries of the foliation structure. 
To be precise, the hyper-surfaces are specified by the scalar field $\Phi$ after fixing the unitary gauge $\Phi=\Phi(t)$. 
The corresponding specified unit normal vector $n_\mu$ is written as 
\begin{align}
   n_\mu= -\frac{\delta^0_\mu}{\sqrt{-g^{00}}} \overset{\text{u}}{=}-\frac{\hat{\nabla}_{\mu} \Phi}{\sqrt{2X}}~,
   \label{normal vector}
\end{align}
where the canonical kinetic term $X$ is evaluated as 
\begin{align}
   X\overset{\text{u}}{=}-\frac{1}{2} g^{00}\dot\Phi_0^2\label{X}~,
\end{align}
in the unitary gauge with the assumption that $\dot\Phi_0 >0$. 
Based on this assumption, in the unitary gauge, the lapse function $N$ is related to the scalar field through $N=1/\sqrt{-g^{00}}$. 
The resulting induced metric $h_{\mu\nu}$ is defined as 
\begin{align}
    h_{\mu\nu}=g_{\mu\nu}+n_{\mu}n_{\nu}~.
\end{align} 
Since $n_{\mu}$, as a unit time-like vector, can be seen as 4-velocity of some observer, the corresponding acceleration of the observer is the Lie derivative of the 4-velocity $n_{\mu}$ along itself
\begin{align}
   a_{\mu} & =\pounds_n n_{\mu}= n^\nu\nabla_\nu n_\mu=\text{D}_{\mu}\ln{N}~,
\end{align}
where the spatial derivative is adapted to the induced metric $h_{\mu\nu}$, e.g. $\text{D}_{\mu}\ln{N}\equiv h^{\mu\prime}_{\mu}\nabla_{\mu\prime}\ln{N}$. 
The expression written in spatial derivative of the logarithm of the lapse function implies that it is always tangent to the hyper-surface and also could be regarded as a function related to the spatial derivatives of  $g^{00}$.
The extrinsic curvature is defined by the Lie derivative of the spatial metric with respect to the normal vector
\begin{align}
   K_{\mu\nu} & =\frac{1}{2}\pounds_{n}h_{\mu\nu} =h_{\mu}^{\rho}\nabla_{\rho}n_{\nu}~.
\end{align}
Geometrically, $K_{\mu\nu}$ portrays the curving of hyper-surfaces of equal time $\Sigma_{t}$ in spacetime. Since the normal vector $n_{\mu}$ is a time-like and future-pointing vector, the extrinsic curvature carries the evolution information about the spatial metric. The second order covariant derivative of the scalar field can be expressed in unitary gauge as 
\begin{align}\label{DDphi1}
    \hat{\nabla}_{\mu}\hat{\nabla}_{\nu}\Phi & =\nabla_{\nu} \nabla_{\mu}\Phi-\nabla_{\rho}\Phi \mathcal{K}_{\phantom{\rho}\mu\nu}^{\rho}\, ,\\
    \nabla_{\mu}\nabla_{\nu}\Phi &\overset{\text{u}}{=}n_{\mu}n_{\nu}\pounds_{n}^{2}\Phi+2n_{(\mu}a_{\nu)}\pounds_{n}\Phi-K_{\mu\nu}\pounds_{n}\Phi~\label{DDphi2}. 
\end{align}
Then $\pounds_{n}^{2}\Phi$ corresponds to the second order time derivatives of scalar field, which would lead to the so-called Ostrogradsky ghost.

A general EFT action involving operators up to the second order in curvature-based geometry is \cite{Gubitosi:2012hu}
\begin{align}
\label{curvEFTact}
 S &= \int d^4x \sqrt{-g} \Big[ \frac{M^2_P}{2} \Psi(t)R - \Lambda(t) - b(t)g^{00} \nonumber \\
 & \ \ \ \ \ \ \ \ \ \ \ \ \ \ \ \ \ \ \ \ \
 + M_2^4(\delta g^{00})^2 -\bar{m}^3_1 \delta g^{00} \delta K -\bar{M}^2_2 \delta K^2 - \bar{M}^2_3 \delta K^{\nu}_{\mu} \delta K^{\mu}_{\nu} \nonumber \\
 &\ \ \ \ \ \ \ \ \ \ \ \ \ \ \ \ \ \ \ \ \
 + m^2_2 h^{\mu\nu}\partial_{\mu} g^{00}\partial_{\nu}g^{00} +\lambda_1\delta R^2 + \lambda_2 \delta R_{\mu\nu}\delta R^{\mu\nu} +\mu^2_1 \delta g^{00} \delta R  \nonumber \\
 &\ \ \ \ \ \ \ \ \ \ \ \ \ \ \ \ \ \ \ \ \
 + \gamma_1 C^{\mu\nu\rho\sigma} C_{\mu\nu\rho\sigma} + \gamma_2 \epsilon^{\mu\nu\rho\sigma} C_{\mu\nu}
^{\quad\kappa\lambda} C_{\rho\sigma\kappa\lambda} \Big] ~,
\end{align}
where the $K$ is the trace of extrinsic curvature and the $C_{\mu\nu\rho\sigma}$ is the Weyl tensor. All the dynamical DoFs are encoded in the metric perturbations. 
In the unitary gauge, the scalar field is used to define the spacelike hypersurfaces and thus absorbed in the geometric quantities. 
Relevant DoFs are described by metric perturbations around a homogeneous and isotropic background. 

Based on our requirements and consideration, 
we first rewrite the scalar field part of the action \eqref{action scalar} in unitary gauge as 
\begin{align}\label{actionscalaru} 
    \mathcal{L}_{\text{scalar}}(g^{\mu\nu};\Phi,\hat\nabla_\mu \Phi, \hat\nabla_{[\mu} \hat\nabla_{\nu]}\Phi)\overset{\text{u}}{=}\mathcal{L}_{\text{scalar}}(g^{\mu\nu};t, g^{00}, T_{\ \mu 0}^{0}, T_{\ \mu\nu}^{0})~,
\end{align}
where $T_{\ \mu 0}^{0}$ and $T_{\ \mu\nu}^{0}$ are spatial vector and tensor, respectively.
In deriving the above, the relations
\begin{align}
(\hat{\nabla}_{\mu}\hat{\nabla}_{0}-\hat{\nabla}_{0}\hat{\nabla}_{\mu})\Phi & \sim  T_{\ \mu 0}^{0}~, \\
   (\hat{\nabla}_{\mu}\hat{\nabla}_{\nu}-\hat{\nabla}_{\nu}\hat{\nabla}_{\mu})\Phi & \sim  T_{\ \mu\nu}^{0}~ 
\end{align}
are applied.
Non-minimal coupling part \eqref{action coupling} becomes
\begin{align}
 &\mathcal{L}_{\text{couple}}(T,T^\mu,T^{\mu \nu \rho} ;\Phi,X,\hat\nabla_\mu \Phi, \hat\nabla_{[\mu} \hat\nabla_{\nu]}\Phi)  \nonumber\\ 
\overset{\text{u}}{=} &\mathcal{L}_{\text{couple}}(T,T^0,T^\mu,T^{0 \nu 0},T^{0 \nu \rho},T^{\mu 0 \rho},T^{\mu \nu \rho} ;t, g^{00}, T_{\ \mu 0}^{0}, T_{\ \mu\nu}^{0})\label{actioncouplingu} ~,
\end{align}
after taking the unitary gauge.
In the EFT form, $\Phi$ and its derivatives would be regarded and absorbed into a time-dependent function. 
Because in this form $T_{\ \mu 0}^{0}$ and $ T_{\ \mu\nu}^{0}$ are introduced via $\hat\nabla_{[\mu} \hat\nabla_{\nu]}\Phi$, coupling terms such as $T^\mu T_{\ \mu 0}^{0}$ are allowed within our framework. 

\subsection{Leading order operators} 

We have already arranged all the basic building blocks in \eqref{actionscalaru} and \eqref{actioncouplingu}. Then we expand them in powers of number of perturbations around an FLRW background. 
It is straightforward to separate a tensor $X$ into its unperturbed (background) part $X^{(0)}$ and its perturbation part $\delta X= X-X^{(0)}$, which allows us to decompose the action into primary operators and other perturbation operators based on their  contributions to the perturbations. 
As a straightforward example, let us consider an operator made by the contraction of two tensors $X$ and $Y$. By expanding $X= X^{(0)}+\delta X$ and $Y= Y^{(0)}+\delta Y$, we have
\begin{align}
\label{TGexpr}
 XY = \delta X \delta Y + X^{(0)}Y + XY^{(0)} - X^{(0)} Y^{(0)} ~.
\end{align}
Note that in an FLRW background one can always express the unperturbed tensors $X^{(0)}$ and $Y^{(0)} $ as functions of $g_{\mu\nu}$, $n_{\mu}$ and $t$. Then we combine them in the EFT form and consider that
\begin{itemize}
    \item all the background quantities can be absorbed in $\Lambda(t)$.
\end{itemize}
Moreover, the definition of these perturbation operators makes their expansion particularly advantageous when investigating the linear and higher order perturbations around an FLRW background. 
Consequently, it is reasonable to focus on operators up to the required order, rather than encompassing the entire action.

In this subsection, we mainly focus on the leading operators like $X^{(0)}Y$.
Therefore, based on $\mathcal{L}_{\text{scalar}}$ \eqref{actionscalaru}, we get that  
\begin{itemize}
     \item $b(t) g^{00}$ is the only possible option and shows clearly the existence of the dynamical scalar degree of freedom.
\end{itemize}
In the case of $f(T)$ gravity, Taylor expansion 
    \begin{align}
    f\left(T\right)=f(T^{(0)})+f_{T}(T^{(0)})\left(T-T^{(0)}\right)+\frac{1}{2}f_{TT}(T^{(0)})\left(T-T^{(0)}\right)^{2}+\cdots
    \end{align}
can be applied to obtain its corresponding operators from pure gravity side, where $T^{(0)}$ is used to represent the torsion scalar at the background level with the value $T^{(0)}=6H^{2}$ and the subscripts “$T$" and “$TT$" refer to first and second order derivative with respect to $T$, respectively. Even though we are dealing with a torsion-based theory, replacing the operator $T$ through the relation \eqref{eq:relationRT} as a combination of  Ricci scalar $R$ and the boundary term $\nabla_\mu T^\mu$ allows for a straightforward return to GR case in the appropriate limit. 
Then, we propose that 
\begin{itemize}
    
    \item $\Psi(t)R$ should be included as a basic operator in the gravity part in order to reduce to the case of GR by taking $\Psi(t)=1$.  
    
\end{itemize}
In this case, the linear term $\nabla_\mu T^\mu$ with coefficient should be transformed into the contracted torsion tensor $T^0$ by integration by parts. 
Moreover, additional appropriate operators are derived from the coupling part $\mathcal{L}_{\text{couple}}$, in which
\begin{itemize}

    \item $d(t)T^{0}$ term should be included with $d(t)$ an arbitrary function of time.
    
\end{itemize}
It is easy to see that the coupling terms such as $T^\mu \nabla_\mu\Phi$ would contribute to the operator $T^{0}$ in the unitary gauge. This implies that the operator $T^{0}$ can enter the Lagrangian in various ways, not only the replacement of the torsion scalar but also the coupling terms. 
Put differently, even in the context of $f(T)$ gravity, there is no necessity to impose any specific relation between $d(t)$ and $b(t)$ concerning these coupling terms. 
This scenario goes beyond the scope of our prior investigation \cite{Hu:2023juh}.
As of now, we obtain the same EFT form at leading order comparing with the form \cite{Li:2018ixg} given by 
\begin{align}
 S = \int d^4x \sqrt{-g} \Big[ \frac{M^2_P}{2} \Psi(t)R - \varLambda(t) - b(t) 
g^{00} + \frac{M^2_P}{2} d(t) T^0 \Big]+ S^{(2)}~,
 \label{generalEFT}
\end{align}
where $\Psi(t)$, $\varLambda(t)$, $b(t)$ and $d(t)$ are time-dependent coefficients.
Some of other leading operators, such as linear $K$ or the boundary term $B$, can be canceled by integration by part. By taking $\Psi(t)=1$, $\Lambda(t)=d(t)=0$ and $b(t)=\frac{1}{2}\dot\Phi^2$, the leading order action \eqref{generalEFT} reduces to GR with a canonical kinetic term of a scalar field. According to the separation form of the EFT action, in which the background and perturbations are systematically addressed through an expansion, all terms exhibiting a leading contribution of second order perturbations are encompassed in $S^{(2)}$. 
In this paper, we aim to investigate linear order perturbation and the resulting strong coupling issue. Therefore, our focus is on operators up to the second order. Higher order operators are disregarded, and $S^{(2)}$ proves adequate for our analysis. 

\subsection{Second-order operators}

At this time, we do not consider the boundary term $B$ in the EFT action. We only consider $f(T)$ gravity from the pure gravity side. We impose the restriction that
\begin{itemize}
    \item  with the definition $\delta T=T-T^{(0)}$, only $\delta T\delta T$ operator would be added in $S^{(2)}$ from the pure gravity side. 
\end{itemize} 
Operators related to Riemannian curvature, such as $\delta R_{\mu\nu}$, $\delta R$ and Weyl tensor then would be ignored in this torsional EFT action. 

Second order derivatives of the scalar field are also necessary to satisfy our specific requirement.
On the one hand, for pure gravity sector, we consider torsion scalar as the only building block. And the theory in linear order is written in terms of the equivalent boundary term $d(t)T^0$ with Ricci scalar $R$, instead of decomposed spatial curvature scalar $\,^3 R$ with other terms involved extrinsic curvature $K_{\mu\nu}$. It means that no such $\delta K^\mu_\nu$ and $\partial_{\mu} g^{00}$ operators are produced from this sector.
On the other hand, operators with $\delta K^\mu_\nu$ and $\partial_{\mu} g^{00}$, namely the perturbations of the $t=const$  hyper-surfaces, can originate from the scalar field side according to the expression of second order derivatives \eqref{DDphi1} and \eqref{DDphi2}. The equations \eqref{X} and \eqref{DDphi2} also imply that $\delta K^\mu_\nu$ and $\partial_{\mu} g^{00}$ are related with second order derivative of scalar field and non-linear higher order derivatives are likely to make the theory suffer from instability. Then we impose further conditions that only pairs of commutators $ \hat\nabla_{[\mu} \hat\nabla_{\nu]}\Phi$ are allowed and thus these higher order operators should not be included in this action, exactly as the scalar and coupling parts we imposed in \eqref{actionscalaru} and \eqref{actioncouplingu}. In summary, from both perspectives operators with $\delta K^\mu_\nu$ and $\partial_{\mu} g^{00}$ would not be included in this action.
Then based on $\mathcal{L}_{\text{scalar}}$ \eqref{actionscalaru}, it is easy to see 
\begin{itemize}

    \item $\delta g^{00}\delta g^{00}$, $\delta T^{0\mu\nu}\delta{T^{0 }}_{\mu \nu}$ and $\delta T_{\ \mu 0}^{0} \delta{T^{0 \mu}}_{0} $ should be added in $S^{(2)}$.
    
\end{itemize} 
It also means that these operators enter in the action by quadratic contraction commutators, instead of the quadratic torsion contraction.
Subsequently, we turn our attention to the coupling terms. Based on the form of \eqref{actioncouplingu}, it is straightforward to get irreducible contractions $\delta T^{\mu}\delta{T^{0 }}_{\mu 0}$,$\delta {T^{\mu 0 \nu}} \delta{T_{\ \mu\nu}^{0}}$ and reducible contractions $\delta T^0 \delta T$, $\delta T^0 \delta T^0 $, $\delta T^0 \delta g^{00}$, $ \delta T \delta g^{00}$, respectively.
All operators are introduced without Ostrogradsky ghost. 

In all, second order operators satisfying all the conditions above are
\begin{align}\label{second order action}
 S^{(2)}= \int d^4x \sqrt{-g} \Big[& \nu_1\delta T\delta T + c_1\delta T\delta T^0 + d_{1} \delta T \delta g^{00} + d_2\delta T^0 \delta T^0 \\ \nonumber
& + f_1 \delta T^0 \delta g^{00} + g_1\delta g^{00}\delta g^{00} + d_3\delta T^{0\mu\nu}\delta{T^{0 }}_{\mu \nu} \\ \nonumber
& + d_4\delta T_{\ \mu 0}^{0} \delta{T^{0 \mu}}_{0} + d_5\delta T^{\mu}\delta{T^{0 }}_{\mu 0} + d_{6} \delta {T^{\mu 0 \nu}} \delta{T_{\ \mu\nu}^{0}}\Big]~,
\end{align}
with coefficients $\nu_1,c_1,d_i,f_1,g_1$, $i=1,\dots,6$ being general functions of $t$.

\section{Linear perturbation of a simple model}\label{sec:4}
In the following, we consider a toy model as a simple example. The model is given by 
\begin{align}
 S = \int d^4x \sqrt{-g} \Big[  & \frac{M^2_P}{2} \Psi(t)R - \varLambda(t) - b(t) 
g^{00} + \frac{M^2_P}{2} d(t) T^0 +\frac{M^2_P}{2} F(t) \delta T\delta T \Big] ~.
 \label{model}
\end{align}
Only one second order operator $\delta T\delta T$ is considered in this toy model.
Furthermore, the equivalent form of $f(T)$ gravity \cite{Wright:2016ayu} after performing a conformal transformation is also included in the model as a special case with coefficients taken as \eqref{conformalfTE coefficients}. We will discuss this conformal equivalent as a special degenerate case in the following with pure  $f(T)$ gravity as well.
In summary, our aim is to investigate if the matter coupling has the potential to alter the inherent constraint structure of gravitational sector, specifically within the framework of $f(T)$ gravity in this study, by perturbation method.
Furthermore, as the first step, our investigation seeks to ascertain the possibility whether such alterations could lead to the reappearance of the scalar mode at the linear order, a feature absent in uncoupled $f(T)$ gravity.
If the propagation of this mode at the linear order is observed, at least we must reconsider the existence of a so-called strong coupling problem in the cosmological background under this scenario.

\subsection{Perturbed tetrads}
As previously mentioned, to address the question of whether the EFT model, which corresponds to the simple coupled $f(T)$ gravity case, exhibits signs of strong coupling, it is necessary to study its linear scalar perturbations around a cosmological background. 
For convenience, we concentrate solely on the scalar part.
Accordingly, the perturbed tetrads take the form as
\begin{equation}       e_{\phantom{(1)A}\mu}^{(1)A}\equiv\left(\begin{array}{cc}
1+\phi & a\partial_{i}\chi\\
\partial^{a}\chi & a\left(1-\psi\right)\delta_{i}^{a}
\end{array}\right) \label{basictetrad}
\end{equation}
in the Newtonian gauge, where $\phi$, $\psi$, $\chi$ are the corresponding scalar perturbations and $a$ is the scale factor. And we have neglected the pseudo-scalar mode to avoid the parity violation. In order to have expressions for the tetrads that are higher order in perturbations, we employ $e_{~\mu}^{A}=\delta_{~\nu}^{A}e_{~\mu}^{\nu}$ and make the exponential ansatz  \cite{Li:2018ixg},
\begin{equation}
e_{\phantom{\nu}\mu}^{\nu}\coloneqq\left(e^{\bm{m}}\right)_{\phantom{\nu}\mu}^{\nu}\equiv\delta_{\phantom{\nu}\mu}^{\nu}+m_{\phantom{\nu}\mu}^{\nu}+\frac{1}{2}m_{\phantom{\nu}\rho}^{\nu}m_{\phantom{\nu}\mu}^{\rho}+\cdots,
\label{Eexpand}
\end{equation}
with the matrix $m_{\phantom{\nu}\mu}^{\nu}$ being linear in perturbations. For our purpose, we choose
    \begin{equation}
        m_{\phantom{\nu}\mu}^{\nu}\coloneqq\bar{e}_{A}^{\phantom{A}\nu}e_{\phantom{(1)A}\mu}^{(1)A}-\delta_{\phantom{\nu}\mu}^{\nu}~,
    \end{equation}
where 
    \begin{equation}
        \bar{e}_{A}^{\phantom{A}\nu}=\left(\begin{array}{cc}
1 & 0\\
0 & \frac{1}{a}\delta_{a}^{i}
\end{array}\right)
    \end{equation}
is the background tetrads, and $e_{\phantom{(1)A}\mu}^{(1)A}$ is the linear order tetrads defined in eq. (\ref{basictetrad}).
Then we get the perturbed tetrads up to the second order as: 
\begin{align}   	  e^0_{\phantom{0}\mu} =& \delta^0_{\mu} \big( 1 +\phi +\frac12\phi^2 
+\frac12\partial_i\chi \partial_i\chi \big) + a\delta^i_{\mu} \Big[ \partial_i 
\chi + \frac12(\phi\partial_i\chi -\psi\partial_i\chi) \Big] ~, \\  e^a_{\phantom{0}\mu} =& 
a\delta^i_{\mu} \delta^a_i \big( 1-\psi+\frac12\psi^2 \big) + \frac{a}{2} 
\delta^i_{\mu} \delta^a_j \partial_i \chi \partial_j \chi + \delta^0_{\mu} 
\delta^a_i \Big[ \partial_i\chi +\frac12(\phi\partial_i\chi - 
\psi\partial_i\chi) \Big] ~, \\  e^{\phantom{0}\mu}_0 =& \delta^{\mu}_0 \big( 1 -\phi 
+\frac12\phi^2 +\frac12\partial_i \chi\partial_i\chi \big) + 
\frac1a\delta_i^{\mu} \Big[ -\partial_i\chi +\frac12(\phi\partial_i\chi - 
\psi\partial_i\chi) \Big] ~, \\  e^{\phantom{0}\mu}_a =& \frac1a \delta_i^{\mu} \delta_a^i 
\big( 1 + \psi + \frac12 \psi^2 \big) + \frac{1}{2a} \delta_i^{\mu} \delta_a^j 
\partial_i \chi \partial_j \chi + \delta_0^{\mu} \delta_a^i \Big[ 
-\partial_i\chi +\frac12(\phi\partial_i\chi - \psi \partial_i \chi) \Big] ~. 
\end{align}

Through St\"{u}ckelberg trick we can restore the full spacetime diffeomorphism invariance of the theory. 
The Nambu-Goldstone theorem is applicable in the scenario of the spontaneous breaking of time reparametrization symmetry.
Following a time coordinate transformation of the form $t\rightarrow t+\pi$ on the EFT action, the Goldstone mode $\pi$ manifests in the action. 
Technically, specific examples that will be employed subsequently are provided.
$T^{(0)}$ (and other background quantities) should be regarded as a time-dependent function and expanded as
\begin{align}
T^{(0)}\rightarrow  T^{(0)}+\dot{T}^{(0)}\pi+\frac12\ddot{T}^{(0)}\pi^2+\cdots~.
\end{align}
The part involving temporal components is subjected to a distinct treatment.
For the time-component of the contracted torsion $T^{0}$, we have
\begin{align}
T^{0}\rightarrow T^{0}+\partial_{\mu}\pi\,T^{\mu}~.
\end{align}
In order to study linear perturbation of this torsional EFT action, St\"{u}ckelberg trick becomes essential to ensure consistent results with a covariant form. Particularly, employing perturbed tetrads fixed to the Newtonian gauge eliminates gauge freedom for scalar perturbations.

At background level, we will get the same background equations given in \cite{Li:2018ixg} as
\begin{align}
 b(t) &= M_{P}^2 \Psi \Big( -\dot{H} - \frac{\ddot{\Psi}}{2\Psi} + \frac{H\dot{\Psi}}{2\Psi} - \frac{\dot{d}}{4\Psi} + \frac{3Hd}{4\Psi} \Big) ~, \label{b(t)} \\
 \Lambda(t) &= M_{P}^2\Psi \Big( 3H^2 + \frac{5H \dot{\Psi}}{2\Psi} + \dot{H} + \frac{\ddot{\Psi}}{2\Psi} + \frac{\dot{d}}{4\Psi} +\frac{3Hd}{4\Psi} \Big) ~.
\end{align}
In this work, we solve the expression of $\dot H$ from the first equation \eqref{b(t)} as 
\begin{equation}\label{H}
\dot H = \frac{-4 b(t)+3 M_p^2 d(t) H-M_p^2 \dot d(t)+2 M_p^2 H \dot \Psi(t)-2 M_p^2 \ddot \Psi(t)}{4 M_p^2 \Psi(t)}~,
\end{equation}
and replace it by this expression.
Throughout this paper, we assume that $\Psi(t)$ would not vanish.

Proceeding to linear-order perturbations. We first perform variations with respect to each variable. 
Through variations with respect to both $\pi$ and $\psi$, we obtain two equations of motion. In this model, which generally involves two dynamical variables, namely, $\pi$ and $\psi$, the non-degeneracy of the $2\times2$ Jacobian formed by the second-order time derivatives in both equations of motion is a crucial necessary condition for our purpose.
Additionally, two constraints from the variation with respect to $\chi$ and $\phi$ are derived as
\begin{align}\label{constraint1}
 &  a\Big[M_p^2\left(-a d(t)(H\pi+\psi)+8F(t) H^2(\partial^2 \chi+3 aH \phi+3 a \dot\psi )\right)\\ \nonumber
&\ \ \ \ -6 a F(t)H^2 \frac{4 b(t)+M_p^2\left(-3 d(t)H+\dot d(t)-2 H \dot\Psi(t)+2 \ddot\Psi(t)\right)}{\Psi(t)}\pi\Big]=0~, 
\end{align}
and
\begin{align} \label{constraint2}
&-2 M_p^2 a \Psi(t)\left(-\partial^2\psi+3 a^2 H(H \phi+\dot\psi)\right) \\\nonumber
&+\frac{3 a^3 \pi}{8 \Psi(t)} \left(d(t)-48F(t) H^3+2 \dot \Psi(t)\right)\left[4 b(t)+M_p^2\left(-3 d(t) H+\dot d(t)-2 H \dot \Psi(t)+2 \ddot \Psi(t)\right)\right] \\\nonumber
& + \frac{1}{2} a\Big\{4 a^2 b(t)\left(3 H \pi+\phi-\dot \pi\right)+48 M_p^2 aF(t) H^3\left(\partial^2\chi+3 a H \phi+3 a \dot\psi\right) \\\nonumber
&\qquad\quad -M_p^2\Big[ d(t)\left(\partial^2 \pi+3 a^2(3 H^2 \pi+2 H \phi-H \dot\pi+\dot\psi)\right)-3 a^2 H \pi \dot d(t)+6 a^2 H^2 \pi \dot \Psi(t) \\\nonumber
& \qquad \qquad\qquad+12 a^2 H \phi \dot \Psi(t)+2 \partial^2 \pi \dot \Psi(t)- 6 a^2 H \dot \pi \dot \Psi(t)+6 a^2 \dot\psi \dot \Psi(t)-6 a^2 H \pi \Big] \Big\}=0~, 
\end{align}
where “$\partial^2$" stands for a contraction of spatially partial derivatives as “$\partial_i\partial^i$". Due to the complexity of the constraints, we will categorize them into distinct cases to discuss the potential existence of two scalar modes propagating in a cosmological background.

\subsection{$F(t)=0$ case}\label{sec:4.2}
First, let us examine the case where $F(t)=0$.
This case usually corresponds to linear $T$ theory rather than a general functional form of $f(T)$ theory. The discussion of this case with additional scalar field is more for overall completeness and subsequent research on the equivalence of theoretical forms.
One can easily reduce the constraint \eqref{constraint1} to
\begin{align}
    d(t)(H\pi+\psi)=0~ \label{constraint1a}~.
\end{align}
In general, $d(t)$ is not equal to zero. One solution is to consider
\begin{equation}
    \psi=-H\pi~,
\end{equation}
implying that $\psi$ remains constrained by $\pi$. 
Employing conventional constraint analysis techniques, we utilize the temporal evolution of this constraint to derive additional constraints. Consequently, there is a reduction in the system's DoFs, even when the initial $2\times2$ Jacobian matrix satisfies the non-degeneracy condition. This further rules out the potential emergence of two dynamical scalar modes in this particular subcase.

The other choice is $d(t)=0$, which means that there are certain coupling terms aimed at canceling the contribution from torsion, given our assumption of torsional geometry. Despite the expression resembling a curvature-based theory, there could be non-minimal coupling with the scalar field, acting as a function of time absorbed in its coefficient function, rather than reducing to exactly minimal coupled GR case. In this case, the constraint \eqref{constraint1a} is automatically satisfied. In other words, we lose a primary constraint after fixing $d(t)=0$. The constraint equation \eqref{constraint2} is also simplified by this conditon. Then we can easily solve $\phi$ as $\phi=\phi(\pi,\psi;\dot \pi, \dot \psi;\partial^2)$. 
Because only time derivatives would be involved and contribute to the degenerate analysis, without loss of generality we only show the relevant part associated with the Jacobian as 
\begin{equation}\label{velocityphi}
\phi=\frac{2 b(t) \dot \pi+3 M_p^2\left[2 \Psi(t) H \dot \psi+\dot \Psi(t) \left(-H \dot \pi+\dot \psi\right) \right]}{2\left[b(t)-3 M_p^2 H\left(\Psi(t) H+\dot \Psi(t)\right)\right]}+\cdots~.
\end{equation}
where the ellipses “$\cdots$" represent for the part of solution which does not affect dynamics. We make use of time evolution of this solution and get the relation of $\dot\phi=\dot\phi(\pi,\psi;\dot \pi, \dot \psi; \ddot \pi, \ddot \psi;\partial^2)$, which is up to second order time derivatives.
Then the relevant second order time derivatives in the equations of motion obtained by the variation of $\pi$ and $\psi$ would be changed from 
\begin{align}
  & -\frac{1}{2}a^{3}\left[4b(t)\ddot{\pi}+3M_{p}^{2}\left(d(t)+2\dot\Psi(t)\right)\ddot{\psi}-4 b(t)\dot \phi+3 M_p^2 H\left(d(t)+2 \dot \Psi(t)\right)\dot \phi\right]+\cdots~,	\\
& -\frac{3}{2}M_{p}^{2}a^{3}\left[\left(d(t)+2\dot\Psi(t)\right)\ddot{\pi}-4\Psi(t)\ddot{\psi}-\left(d(t)+4 \Psi(t) H+2 \dot\Psi(t)\right)\dot \phi\right]	+\cdots~,
\end{align}
into
\begin{align}
 &\frac{3M_{p}^{2}a^{3}}{4\left[b(t)-3M_{p}^{2}H\left(\Psi(t)H+\dot\Psi(t)\right)\right]}\left(H\ddot{\pi}+\ddot{\psi}\right)\times \\ \nonumber
&\ \ \ \  \left[-2b(t)\Big(d(t)-4\Psi(t)H\Big)+3M_{p}^{2}H\dot\Psi(t)\left(d(t)+2\dot\Psi(t)\right)\right] +\cdots~,\\
&\frac{3M_{p}^{2}a^{3}}{4\left[b(t)-3M_{p}^{2}H\left(\Psi(t)H+\dot\Psi(t)\right)\right]}\left(H\ddot{\pi}+\ddot{\psi}\right)\times \\ \nonumber
&\ \ \ \  \left\{8b(t)\Psi(t)+3M_{p}^{2}\left[2\dot\Psi(t)^{2}+d(t)\left(2\Psi(t)H+\dot\Psi(t)\right)\right]\right\}+\cdots~.
\end{align}
Upon replacing all of $\phi$ with its solution in the perturbation action, only the specific combination $H\ddot\pi+\ddot\psi$ is shown. 
Although there are two dynamical variables appear in the equations of motion, one can easily reduce the number through a redefinition of variables and get a simplified action involving only a single dynamical variable. 
The Jacobian of these higher order terms with $\ddot \pi$ and $\ddot \psi$ is accordingly generalized after including the contribution of $\dot\phi$ and automatically degenerate.  Subsequently, additional constraints, which is generally expressed as $C(\pi,\psi ;\dot \pi, \dot \psi; \partial^2)=0$, are deduced from a linear combination of these second order equations of motion, where $C$ represents a functional form of these quantities. It results in a direct reduction in the number of dynamical DoFs\footnote{In the appendix A of \cite{Hu:2021yaq}, a thorough analysis of a classical mechanics system with one dynamical and two auxiliary variables is given as a supplementary example. However, we would not follow the steps or classify all cases according to various constraint conditions here.}. Due to the complexity of the form, we do not provide it here. 
If the theory has more constraints, the number of dynamical DoFs will be further reduced. Here, for our purposes, after finding this additional constraint, we stop at this step and do not proceed with further discussion.
Up to this point, both subcases result in a reduction in the rank of the Jacobian. This seems to be unattainable to obtain two propagating scalar modes at linear order within this category $F(t)=0$ of the toy model.

Additionally, we would like to include a point that is not pertinent in this manuscript. 
The denominator $b(t)-3 M_p^2 H\left(\Psi(t) H+\dot \Psi(t)\right)$ above depends on $H$ as long as $\Psi(t)\neq0$. 
In some special forms, $H$ can be regarded as a resulting background value of some geometry quantity, such as $T^{(0)}$.
This renders it a differential equation, and upon rewriting the relation, it can be solved as a specific function dependent on this geometric quantity. However, note that we do not delve into its specific function form in the present discussion.

\subsection{$F(t)\neq0$ case}
We next analyze another scenario that complements it, namely $F(t)\neq0$ case.
With the assumption that $F(t)$ would not vanish, we solve both constraints \eqref{constraint1}, \eqref{constraint2} and express $\partial^2\chi$ and $\phi$ with the assumption 
\begin{equation}
b(t) \neq \frac{3}{2} M_p^2 H\left(d(t)+2 \Psi(t) H+2 \dot\Psi(t)\right)~.
\end{equation}
For the same reason mentioned above, we only list time derivatives in both solutions as 
\begin{align}
   \partial^2\chi &= 3 a \frac{-4 b(t)+3 M_p^2 H\left(d(t)+2 \dot\Psi(t)\right)}{4 b(t)-6 M_p^2 H\left[d(t)+2\left(\Psi(t) H+\dot\Psi(t)\right)\right]}\left(H \dot \pi+\dot \psi\right)+\cdots~,
   \label{timechi}
\end{align}
and
\begin{align}
    \phi & = 3 M_p^2 \frac{ d(t)+4 \Psi(t) H+2 \dot\Psi(t)}{4 b(t)-6 M_p^2 H\left(d(t)+2\Psi(t) H+2\dot\Psi(t)\right)}\dot \psi \nonumber \\ 
    & \ \ \ \ \quad + \frac{4 b(t)-3 M_p^2 H\left(d(t)+2 \dot \Psi(t)\right)}{4 b(t)-6 M_p^2 H\left(d(t)+2\Psi(t) H+2\dot\Psi(t)\right)}\dot \pi+\cdots~.
\label{timephi}
\end{align}
In this case, both solutions of non-dynamical variables $\phi$ and $\chi$ involve “velocity" $\dot \pi$ and $\dot \psi$.
By making use of both solutions, second order time derivatives in both equations of motion, which take the form as 
\begin{equation}
\frac{1}{2} a^3\Big[4 b(t)(\dot \phi-\ddot \pi)-3 M_p^2(H \dot \phi+\ddot \psi)\left(d(t)+2 \dot\Psi(t)\right)\Big]~,
\end{equation}
and 
\begin{align}
& -\frac{3}{2} M_p^2 a^2\Big[32 F(t) H^2\left(\partial^2 \dot \chi+3 a(H \dot \phi+\ddot \psi)\right)\nonumber\\
& \ \ \ \  -a\left(d(t)(\dot \phi-\ddot \pi)+4 \Psi(t)(H \dot \phi+\ddot \psi)+2(\dot \phi-\ddot \pi) \dot\Psi(t)\right)\Big]~,
\end{align}
are expanded as
\begin{align}
& \frac{3 M_p^2 a^3 \left[3 M_p^2 H \dot\Psi(t)\left(d(t)+2 \dot\Psi(t)\right)-2 b(t)\left(d(t)-4 \Psi(t) H\right) \right]}{4\left[b(t)-3 M_p^2 H\left(\Psi(t) H+\dot\Psi(t)\right)\right]} \left(H\ddot \pi +\ddot \psi \right)~,
\end{align}
and
\begin{align}
& -3 M_p^2 a^3 H \Big\{-16 b(t)^2 \Psi(t)+d(t)^2\Big(2 \Psi(t) H+\dot\Psi(t)\Big)\\ \nonumber
&\ \ \ \ +6 M_p^2 b(t)\Big[8 \Psi(t)^2 H^2+8 \Psi(t) H \dot\Psi(t)-2 \dot\Psi(t)^2 \\ \nonumber
&\ \ \ \ +d(t)\Big(2 \Psi(t) H-48 F(t) H^3-\dot\Psi(t)\Big)\Big]+9 M_p^4 H\Big[4 \dot\Psi(t)^2\Big(\Psi(t) H+\dot\Psi(t)\Big) \\ \nonumber
&\ \ \ \ +d(t)\Big(4 \Psi(t)^2 H^2-96 \Psi(t) F(t) H^4+6 \Psi(t) H \dot\Psi(t)+4 \dot\Psi(t)^2\Big)\Big]\Big\}\left(H\ddot \pi +\ddot \psi \right)\bigg/ \\ \nonumber
& \Big\{4\left[b(t)-3 M_p^2 H\left(\Psi(t) H+\dot\Psi(t)\right)\right]\left[ 2 b(t)-3 M_p^2 H\left(d(t)+2\Psi(t) H+2\dot\Psi(t)\right)\right]\Big\}~.
\end{align}
It is easy to see that the dynamical variables $\pi$ and $\psi$ enter in a specific combination form $H\ddot\pi+\ddot\psi$ among both equations of motion.
And the Jacobian automatically degenerates when considering time evolution of both solutions of $\phi$ and $\chi$ above. 
Further constraints would be introduced by this degeneracy and imply that at most one dynamical scalar mode is present in this model without any further conditions.

\subsection{Primary degenerate cases}
The variation with respect to $\pi$ and $\psi$ reveals respectively the parts of second order time derivatives in their equations of motion:
\begin{equation}
-\frac{1}{2} a^3\Big[3 M_p^2\left(d(t)+2 \dot\Psi(t)\right)\ddot \psi+ 4 b(t)\ddot \pi\Big]~,
\end{equation}
and 
\begin{equation}
 -\frac{3}{2} M_p^2 a^3\Big[4 \Big(12 F(t) H^3- \Psi(t)\Big) \ddot \psi+\left(2\dot\Psi(t)+d(t)\right) \ddot \pi \Big]~.
\end{equation}
At this level, further constraints can be derived easily in this primary degenerated Jacobian, which satisfy 
\begin{equation}\label{Degenerateb}
b(t) = -\frac{3 M_p^2\left(d(t)+2 \dot \Psi(t)\right)^2}{16\Big(\Psi(t)-12 F(t) H^2\Big)}~,
\end{equation}
or 
\begin{align}\label{DegeneratePsi}
    \Psi(t)=12F(t)H^2~,\quad \dot \Psi(t)=-\frac{d(t)}{2}~.
\end{align}
Then we recognize the significance of coupling terms. Arbitrary functions $d(t)$ and $\Psi(t)$ prevent the theory from acquiring additional primary constraint. 
And pure $f(T)$ gravity theory falls into the first case, i.e. \eqref{Degenerateb}, with $b(t)=0$ when we take
\begin{align}
   \dot \Psi(t)=-\dot f_T=-2F(t)~, \quad d(t)=2\dot f_T=4\dot F(t)~.
\end{align}
In this case, the coefficient functions $\Psi(t)$, $F(t)$ and $ d(t) $ should satisfy specific requirements according to operator expansion.
Only $\pi$ is a dynamical variable and $\psi$ is fixed as $\psi=-H \pi$.
Additionally, $\chi$ and $\phi$ would be solved and expressed in terms of $\pi$ as 
\begin{align}
\chi & =\frac{\pi}{a}~, \\ 
\label{fTphi}
\phi & =\dot{\pi}-\frac{\partial^{2}\pi}{3a^{2}H}. 
\end{align}
Considering time evolution of \eqref{fTphi}, new constraint would be obtained, indicating that no scalar mode will be propagating in the linear perturbation level. Then we obtain consistent result of $f(T)$ gravity by this toy model as we expected. 

Another interesting degenerate case is the conformal $f(T)$ equivalent.
Following the same strategy as in the case of $f(R)$ gravity, a conformal transformation is applied to $f(T)$ gravity. After introducing auxiliary field $\Phi$, the equivalent $f(T)$ action \cite{Wright:2016ayu} can be written into 
\begin{align}
    \label{conformalfTE}
    S_{\text{conform}}&=\frac{1}{16\pi G}\int\,\left[-\hat{T}+\frac{\Phi}{\sqrt{3}}\hat{B}+\frac{1}{2}g^{\mu\nu}\nabla_{\mu}\Phi\nabla_{\nu}\Phi-U(\Phi)\right]\hat{e}\,d^{4}x \nonumber\\ 
    &=\frac{1}{16\pi G}\int\,\left[-\hat{T}-\frac{2}{\sqrt{3}}\hat{T}^{\mu}\nabla_{\mu}\Phi+\frac{1}{2}g^{\mu\nu}\nabla_{\mu}\Phi\nabla_{\nu}\Phi-U(\Phi)\right]\hat{e}\,d^{4}x~,
\end{align}
where the hatted quantities refer to conformal transformed ones.
The presence of the scalar field with an unusual coupling $\hat{T}^{\mu}\nabla_{\mu}\Phi$ contributes to the operator $T^0$ in the EFT action.
Notably, the kinetic term for the conformal mode exhibits an wrong sign. We rewrite the Lagrangian density \eqref{conformalfTE} in our notation and show its form in unitary gauge as
\begin{align}
   \mathcal{L}_{\text{conform}}\bigg|_{\frac{1}{8\pi G}=M_P^2} & =\frac{M_{P}^{2}}{2}\bigg[\hat{R}-\frac{2}{\sqrt{3}}\hat{T}^{\mu}\nabla_{\mu}\Phi+\frac{1}{2}g^{\mu\nu}\nabla_{\mu}\Phi\nabla_{\nu}\Phi-U(\Phi)\bigg]\notag\\
   & \overset{\text{u}}{=}\frac{M_{P}^{2}}{2}\hat{R}-\frac{M_{P}^{2}}{\sqrt{3}}\dot \Phi_{0}\hat{T}^{0}+\frac{M_{P}^{2}}{4}\dot \Phi_{0}^2g^{00}-\frac{M_{P}^{2}}{2}U(\Phi)~.
\end{align}
It is easy to see that the equivalent EFT form of \eqref{conformalfTE} is degenerated with its coefficients
\begin{align}\label{conformalfTE coefficients}
    \Psi(t)=1~,\quad b(t)=-\frac{M_{P}^{2}}{4}\dot \Phi_{0}^2~,\quad d(t)=-\frac{2}{\sqrt{3}}\dot \Phi_{0}~,\quad\dot \Psi(t)=F(t)=0~, 
\end{align}
satisfying degenerate condition \eqref{Degenerateb}.
Further constraint would be derived by this primary degeneracy as $C_1(\pi,\psi,\chi,\phi; \dot \pi,\dot \psi;\partial^2)= 0$, which can be expressed as
\begin{align} \label{constraintconformal}
& \frac{9}{8} M_p^2 a^3 d(t)\Big(d(t)+4 H\Big)\Big(d(t) \dot\pi-4 \dot\psi\Big)-\frac{1}{32} M_p^2 a d(t)\Big\{27 a^2 d(t)^3 \pi \\ \nonumber
& \ \ \ \ +36 a^2 d(t)^2(6 H \pi+\phi)+16\Big[8 a H\partial^2 \chi-4 \partial^2 \psi+9 a^2 H\Big(4 H \phi- \dot d(t)\pi\Big)\Big] \\ \nonumber
&\ \ \ \ +4 d(t)\Big[4\partial^2 \pi+a\Big(36 a H(3 H \pi+2 \phi)+8 \partial^2 \chi-9 a\dot d(t)\pi\Big)\Big]\Big\}=0~.
\end{align}
Let's consider the other two primary constraints \eqref{constraint1} and \eqref{constraint2}. Since $F(t)=0$ with $d(t)\neq0$ would make \eqref{constraint1} lead to $\psi=-H \pi$, both rest primary constraints \eqref{constraint2} and \eqref{constraintconformal} would be simplified after taking these conditions into consideration.
Then $\chi$ and $\phi$ can be solved and expressed by $\pi$ as
\begin{align}
 \chi&= \frac{\pi}{a}~,\\
 \phi&= \dot \pi-\frac{4\partial^2\pi}{3a^2\Big(d(t)+4 H\Big)}~.
\end{align}
with the assumption that the denominator $d(t)+4 H$ does not vanish. Additional constrains $C_2(\pi;\dot\pi;\partial^2)=\partial^4 \pi=0$ is obtained by linear combinations of these equations of motion and constraints with their time evolution. Then the results suggest that no propagating scalar mode exists in linear perturbation.

It is unsurprising that $f(T)$ theory and its conformal equivalent belong to the same category classified by degenerate condition \eqref{Degenerateb}. Based on the form of \eqref{Degenerateb}, as long as the coupling between the gravitational sector $T$ and the scalar field is minimal, namely $\Psi(t)=1$, the scalar field would appear with wrong sign in the kinetic term. However, this ghost mode would not propagate by these constraints. The phenomenon of a conformal transformation generating a phantom scalar field, is very similar to the situation in $f(Q)$ theory \cite{Hu:2023gui}.

\section{Conclusions}\label{sec:5}

It has been demonstrated in Ref.~\cite{Hu:2023juh} that, the scalar DoF of $f(T)$ gravity does not present up to second order scalar perturbations within a flat FLRW background.
Nonetheless, a simple energy scale estimation is provided as well to underscore the credibility of perturbation analysis, since the strong coupling scale significantly exceeds the EFT cut-off.
This study aims to revisit this problem in matter coupling case by incorporating a non-minimally coupled scalar field. 
Specifically, we expect this vanishing scalar mode become dynamical in the flat FLRW background at linear level, implying the simultaneous presence of both dynamical scalar modes in linear perturbation.
With our specific ghost-free action form, we argue that one of these two modes originates from the gravitational sector. 
Consequently, the manifestation of this dynamical scalar mode at the linear order directly challenges the previously established strong coupling criterion of $f(T)$ gravity.
Upon achieving this goal, it will prompt a reassessment of the strong coupling nature of gravity theories within a broader context.

In this work, firstly, we begin by formulating the action that describes $f(T)$ gravity coupled with an extra scalar field by adding pure scalar part \eqref{actionscalaru} and coupling part \eqref{actioncouplingu}.
In the abstract Lagrangian considered in this work, second-order derivatives appear only in terms of pairs of commutators to be safe from Ostrogradsky ghost. However, it is straightforward to generalize it through invertible field transformations in practice. Theoretically, two forms connected by invertible transformations are generally dynamically equivalent. 
In brief, the ghost-free model given in this paper can be directly used as a generator to produce a extended healthy counterpart.

The primary focus of this paper is to explore whether the strong coupling problem in $f(T)$ can be solved or alleviated by considering a non-minimal coupled scalar field. 
In this work, we would like to propose this possibility and a  class of action with various coupling terms for further investigation. 
Utilizing the covariant formulation proposed in this study, we rewrite it by EFT approach up to second order. In specific, the leading order operators are given in \eqref{generalEFT} and the second order operators are summarized in \eqref{second order action}. 
As an illustrative example, we introduce a toy model \eqref{model} and investigate its linear scalar perturbations around a cosmological background by the EFT method. 
Moreover, when specific coefficients are chosen, this simple model degenerates into pure $f(T)$ theory and its conformally equivalent form \eqref{conformalfTE}. The consistency of the results simultaneously validates the effectiveness of our approach.

In summary, it is crucial to emphasize that, for this simple model, although the degeneracy of the original two-dimensional Jacobian matrix can be altered by introducing matter coupling, the two constraints derived by variation continue to yield further constraints. 
This indicates the necessity for additional coupling terms to comprehensively explore the corresponding conditions required to prevent the emergence of new constraints.
Then the present dynamical scalar mode of $f(T)$ gravity necessitates a reconsideration of the strong coupling issue that arises in gravitational theories, especially in the presence of matter coupling. 
Once the scalar mode becomes dynamical, further analyses of the strong-coupled energy scale can be conducted without assuming additional parameter relations.
Besides, the EFT approach gives an unified and model-insensitive form where different theories may share a common EFT formulation in a simple form. 
Therefore, the conclusions usually have a broader implications at the perturbation level.
Overall, we believe that taking richer coupling structures into consideration does help to solve the strong coupling problem, since we will have more free coefficient functions to adjust the structure of the constraint system. 
This conclusion, however, requires further substantiation through future research.

\acknowledgments
We thank Martin Krssak, Xin Ren and Emmanuel Saridakis for valuable discussions. 
This work is supported in part by the National Key R\&D Program of China (2021YFC2203100), CAS Young Interdisciplinary Innovation Team (JCTD-2022-20), NSFC (12261131497, 11975020, 12005309), 111 Project for ``Observational and Theoretical Research on Dark Matter and Dark Energy'' (B23042), Fundamental Research Funds for Central Universities, CSC Innovation Talent Funds, USTC Fellowship for International Cooperation, USTC Research Funds of the Double First-Class Initiative. 
We acknowledge the use of computing facilities of SYSU, as well as the clusters LINDA \& JUDY of the particle cosmology group at USTC.

\appendix

\bibliography{reference.bib}
\bibliographystyle{JHEP}

\end{sloppypar}

\end{document}